\documentclass[12pt,a4paper,reqno]{amsart}
\usepackage[utf8]{inputenc}
\usepackage[a4paper,margin=2.5cm]{geometry}
\usepackage{amsmath,amssymb,amsthm,mathtools,abraces}
\usepackage{stmaryrd}
\usepackage{color,xcolor}
\usepackage[hypertexnames=false,hyperfootnotes=false,colorlinks=true,linkcolor=blue,%
citecolor=purple,filecolor=magenta,urlcolor=cyan,unicode,linktocpage=true,pagebackref=false]{hyperref}
\usepackage{yhmath}
\usepackage{verbatim}
\usepackage{enumitem} 
\usepackage{bbm}
\usepackage{csquotes}

\newcommand\be{\begin{equation}}
\newcommand\ee{\end{equation}}

\newcommand\p{\partial}

\newcommand\Tr{{\rm Tr}\,}
\newcommand\diag{{\rm diag}\,}

\DeclareMathOperator{\GL}{GL}

\DeclareMathOperator{\gl}{\mathfrak{gl}}

\makeatletter
\@namedef{subjclassname@2020}{%
  \textup{2020} Mathematics Subject Classification}
\makeatother

\newtheorem*{theorem*}{Theorem}

\theoremstyle{definition}

\usepackage{filecontents}
\title[On W-operators and superintegrability for dessins d'enfant]{
On W-operators and superintegrability for dessins d'enfant}

\author{Alexander Alexandrov}

\address{Center for Geometry and Physics, Institute for Basic Science (IBS), Pohang 37673, Korea
}

\email{ {\tt alexandrovsash at gmail.com}}

\date{\today}

\begin{document}

\begin{abstract} 

In this short note we identify a family of partition functions recently introduced by Wang, Liu, Zhang, and Zhao
with certain specializations of the generating function for dessins d'enfant. This provides a new W-description for orbifold strongly monotone Hurwitz numbers
and new examples of superintegrability in matrix models. 

\end{abstract}

\maketitle


%\tableofcontents

%\bigskip
%\newpage 

%\tableofcontents

\def\thefootnote{\arabic{footnote}}

%%%%%%%%%%%%%%%%%%%%%%%%%%%%%%%%%%%%%%%%%%%%%%%%%%%%%%%%

\section{Introduction}
%\addcontentsline{toc}{section}{Introduction}
%\def\theequation{\arabic{equation}}
\setcounter{equation}{0}

Recently Wang, Liu, Zhang, and Zhao \cite{WLZZ} described a family of the W-operators with a relatively simple action on the Schur functions. These operators generate the partition functions with nice Schur function expansions.

In this short note we identify these partition functions with specifications of the partition function for double strictly monotone Hurwitz numbers (or, equivalently, dessins d'enfant). This identification allows us to construct a double matrix integral representation for these partition functions. Moreover, this family provides a new example of the superintegrability phenomenon in matrix models (actually, two examples associated with two different types of matrix models). Therefore, for the considered family of tau-functions we explicitly describe a set of complimentary descriptions, which include matrix models, superintegrability, W-opearators, and enumerative geometry interpretation. 

The $\beta$-deformed case is a straightforward generalization and will be considered elsewhere.

%%%%%%%%%%%%%%%%%%%%%%%%%%%%%%%%%%%%%%%%%%%%%%%
\section{W-operators} 

Let us describe a family of the partition functions recently introduced by Wang, Liu, Zhang, and Zhao \cite{WLZZ}. These partition functions are described by the operators $\widehat{W}_{-n}$, constructed recursively. 

Consider
\be
\widehat{W}_0=\frac{1}{2}\sum_{k,m\geq  1} \left( k m t_k t_m \frac{\p}{\p t_{k+m}}+(k+m)t_{k+m}\frac{\p^2}{\p t_k \p t_m}\right)+N \sum_{k\geq 1} k t_k \frac{\p}{\p t_k},
\ee
an $N$-deformation of the cut-and-join operator for the simple Hurwitz numbers. With the operator
\be
\widehat{E}_1=\sum_{k\geq 1} (k+1) t_{k+1} \frac{\p}{\p t_k}+N t_1
\ee
we construct
\begin{multline}
\widehat{W}_{-1}=\left[\widehat{W}_0,\widehat{E}_1\right]
=\sum_{k,m\geq  1} \left( k m t_k t_m \frac{\p}{\p t_{k+m-1}}+(k+m+1)t_{k+m+1}\frac{\p^2}{\p t_k \p t_m}\right)\\+2N\sum_{k\geq 1} (k+1)t_{k+1}\frac{\p}{\p t_k}+N^2t_1
\end{multline}
and 
\be
\widehat{W}_{-2}=\left[\widehat{W}_{-1},\widehat{E}_1\right].
\ee

The operators $\widehat{W}_n$ for negative $n<-2$ are defined recursively
\be\label{RR}
\widehat{W}_{-n-1}=\frac{1}{n}\left[\widehat{W}_{-1},\widehat{W}_{-n}\right].
\ee

Operators $\widehat{W}_{-n}$ are homogeneous with $\deg \widehat{W}_{-n}=n$ if one puts $\deg t_k=-\deg \frac{\p}{\p t_k}=k$. Let us stress that all these operators belong to $\gl(\infty)$ algebra of symmetries of the KP hierarchy because $\widehat{W}_{0}$ and $\widehat{E}_{0}$ obviously do.
With these operators we associate the partition functions
\be\label{Zn}
Z_{-n}({\bf t})=\exp\left(\frac{1}{n}\widehat{W}_{-n}\right)\cdot 1.
\ee

The action of the operators $\widehat{W}_{-n}$ on the Schur functions was investigated in \cite{WLZZ}. This allowed the authors to show that 
the generating functions $Z_{-n}$ have a very simple Schur function expansion, namely for $n\geq 2$
\be\label{Zd}
Z_{-n}({\bf t})=\sum_{\lambda}\frac{s_\lambda({\bf 1})}{s_\lambda(\delta_{k,1})}s_\lambda(\delta_{k,n})s_\lambda({\bf t}).
\ee
Here we denote $s_\lambda(\delta_{k,n})=s_\lambda({\bf t})\Big|_{t_k=\delta_{k,n}/n}$ and $s_\lambda({\bf 1})=s_\lambda({\bf t})\Big|_{t_k=N/k}$, so that
\be
\frac{s_\lambda({\bf 1})}{s_\lambda(\delta_{k,1})}=\prod_{i,j \in \lambda} (N+i-j).
\ee

For $n=2$ the function \eqref{Zn} coincides with the partition function of the Gaussian Hermitian matrix model, and the description in terms of the operator $\widehat{W}_{-2}$ is equivalent to the description originally obtained by Morozov and Shakirov \cite{MSh}. In the next section  we will describe matrix models for all $Z_{-n}$.

%%%%%%%%%%%%%%%%%%%%%%%%%%%

\section{Matrix model and superintegrability}

Let us consider a partition function, parametrized by two infinite sets of variables, $t_k$ and $s_k$,
\be\label{Z2}
Z({\bf t},{\bf s})=\sum_{\lambda}\frac{s_\lambda({\bf 1})}{s_\lambda(\delta_{k,1})}s_\lambda({\bf s})s_\lambda({\bf t}).
\ee
It belongs the hypergeometric family of 2-component KP tau-functions, which can be easily deformed to a family of hypergeometric tau-functions of the 2d Toda lattice hierarchy \cite{Kharchev, OS}. This family has a very interesting enumerative geometry interpretation, namely, the hypergeometric tau-functions of 2-component KP hierarchy are the generating functions of the weighted Hurwitz numbers \cite{AMMN0,John}. The tau-function \eqref{Z2} is the generating function for the dessins d'enfant (or double strictly monotone Hurwitz numbers, or hypermaps, or bipartite maps) and is well studied. 

All partition functions $Z_{-n}$ for $n\geq 2$ are given by the specializations of this partition function
\be
Z_{-n}({\bf t})=Z({\bf t},{\bf s})\Big|_{s_k=\delta_{k,n}/n}.
\ee

These specifications of the partition function \eqref{Z2} generate strictly monotone orbifold Hurwitz numbers, see e.g. \cite{ALS} and references therein. Therefore, {\em formula \eqref{Zn} provides a W-representation for the strictly monotone orbifold Hurwitz numbers}. Moreover, these W-operators belongs to the $\gl(\infty)$ algebra, so these formulas describe the $\GL(\infty)$ group elements for these Hurwitz numbers. Thus, the recursion relations \eqref{RR} allows one to describe the point of the Sato Grassmannian and the Kac-Schwarz operators, associated with the tau-functions $Z_{-n}$. The quantum spectral curves for this family were derived in \cite{Do}, their semi-classical limit coincides with the classical spectral curves, recently discussed by Mironov and Morozov \cite{MMN}.

The tau-function \eqref{Z2} can be dewscribed by the two-matrix model (see, e.g., \cite{AMMN}):
\be\label{2MM}
Z({\bf t},{\bf s})=\int_{{\mathcal H}_N}[d \Phi_1]\int_{{\mathcal H}_N}[d \Phi_2]\exp\left(\Tr\left(\Phi_1\Phi_2+\sum_{k=1}^\infty t_k \Phi_1^k+s_k \Phi_2^k\right)\right).
\ee
Here one integrates over the space of Hermitian (to be precise, normal matrices with properly chosen contours) $N\times N$ matrices with the properly normalized flat measures. 

Therefore, we arrive at the matrix model
\be
Z_{-n}({\bf t})=\int_{{\mathcal H}_N}[d \Phi_1]\int_{{\mathcal H}_N}[d \Phi_2]\exp\left(\Tr\left( \Phi_1\Phi_2+\frac{1}{n}\Phi_2^n +\sum_{k=1}^\infty t_k \Phi_1^k\right)\right).
\ee
We expect that the W-description \eqref{Zn} can be derived from the Virasoro and W-constraints for this matrix integral.

For any function $f(X)$ of $N\times N$ matrix let us denote
\be
\left<f \right>_{{\bf s}}= \int_{{\mathcal H}_N}[d \Phi_1]\int_{{\mathcal H}_N}[d \Phi_2] \,f(\Phi_1) \,\exp\left(\Tr\left(\Phi_1\Phi_2+\sum_{k=1}^\infty s_k \Phi_2^k\right)\right).
\ee
Then from the comparison of \eqref{Z2} and \eqref{2MM} and application of the Cauchy formula we conclude
\be
\left<s_\lambda \right>_{{\bf s}}= \frac{s_\lambda({\bf 1})}{s_\lambda(\delta_{k,1})}s_\lambda({\bf s}).
\ee
This provides us a new example of {\em superintegrability}, a nice property of matrix models recently found to be surprisingly universal \cite{SS}.

For particular specifications, associated with the partition functions \eqref{Zn}, we have
\be
\left<s_\lambda \right>_n=\frac{s_\lambda({\bf 1})}{s_\lambda(\delta_{k,1})}s_\lambda(\delta_{k,n}),
\ee
where
\be
\left<f \right>_{n}= \int_{{\mathcal H}_N}[d \Phi_1]\int_{{\mathcal H}_N}[d \Phi_2] \,f(\Phi_1)\, \exp\left(\Tr\left( \Phi_1\Phi_2+ \frac{1}{n} \Phi_2^n\right)\right).
\ee

For $n=2$ the integral with respect to $\Phi_2$ is Gaussian and can be taken explicitly. This way one arrives at the Gaussian Hermitian matrix model description of $Z_2$, obtained in \cite{WLZZ},
\be
Z_{-2}({\bf t})=\int_{{\mathcal H}_N}[d \Phi_1]\exp\left(\Tr\left(-\frac{1}{2}\Phi_1^2 +\sum_{k=1}^\infty t_k \Phi_1^k\right)\right).
\ee
For $n>2$ integral over $\Phi_2$ can also be taken explicitly with the result given in terms of the higher Airy functions.

Another matrix model description for the partition function \eqref{Z2} was derived by Ambj\o{}rn and Chekhov \cite{AC}. Namely, for any function  $f({\bf t})$ we consider the Miwa parametrization
\be
f([\Lambda^{-1}])=f({\bf t})\Big|_{t_k=\frac{1}{k}\Tr \Lambda^{-k}},
\ee
where $\Lambda=\diag(\lambda_1,\dots,\lambda_M)$ is a diagonal matrix. Then \cite{AC}
\be
Z({\bf t},[\Lambda^{-1}])=(-\det \Lambda)^N \int_{{\mathcal H}_M^+}[d \Phi]\exp\left(\Tr \left((N-M)\log \Phi-\Lambda \Phi+\sum_{k\geq 1}t_k \Phi^k\right)\right).
\ee
Here the size of the integration matrix $M$ is independent of the parameter $N$ and one integrates over the space of  positive defined Hermitian matrices. 

Again, from the comparison with the expansion \eqref{Z2} we have the following superintegrability property
\be
(-\det \Lambda)^N \int_{{\mathcal H}_M^+}[d \Phi]\,s_\lambda({\Phi}) \,\exp\left(\Tr \left((N-M)\log \Phi-\Lambda \Phi\right)\right)= \frac{s_\lambda({\bf 1})}{s_\lambda(\delta_{k,1})}s_\lambda([\Lambda^{-1}]).
\ee
Using other matrix models for the families of hypergeometric tau-functions, investigated in particular in \cite{AMMN,MMRP}, one can describe more examples of superintegrability in multi-matrix models associated with Hurwitz numbers.

\subsection*{Acknowledgements}
This work  was supported by the Institute for Basic Science (IBS-R003-D1). 

\bibliographystyle{unsrt}
\bibliography{KPTRref}
\end{document}